\magnification=\magstep1 
\hsize=16.5truecm  
\vsize=24truecm 
\hfuzz=2pt 
\vfuzz=4pt
\pretolerance=5000 
\tolerance=5000 
\parskip=2pt plus 1pt minus 1pt 
\parindent=16pt 
\nopagenumbers
\font\fourteenbf=cmbx10 scaled 1400        

{\vglue 0truecm{
               \baselineskip=20 truept
               \pretolerance=10000
               \raggedright
               \parskip=0pt
               \noindent
               \fourteenbf {A RELATIVISTIC INVARIANT SCHEME FOR THE QUANTUM KLEIN-GORDON AND
DIRAC FIELDS ON THE LATTICE} \vskip 1truecm}

{\pretolerance=10000
 \raggedright
 \noindent {\bf Miguel Lorente}. {\it Departamento de F{\'\i}sica, Universidad de
Oviedo, 33007 Oviedo, Spain}}

\vskip 0.5truecm  {\noindent {\bf Abstract.} We explore the connection between the symmetry
transformations and conservation laws for the Klein-Gordon and Dirac fields on the lattice. The
generators of the space time translations and Lorentz boost (defined on the lattice) are constants
of motion and satisfy the standard commutation relations.} 

\vskip 1truecm  {\noindent \bf 1. Quantization of the Klein-Gordon field}

\medskip We introduce the method of finite differences for the Klein-Gordon scalar field. A
scheme for the wave equation consistent with the continuous case (the truncation error is of
second orden with respect to space and time variables) can be constructed as follows:
$$({1 \over {\tau}^2} \nabla_n \triangle_n \tilde\nabla_j \tilde\triangle_j - 
   {1 \over {\varepsilon}^2} \nabla_j \triangle_j \tilde\nabla_n \tilde\triangle_n +
   M^2 \tilde\nabla_n\tilde\triangle_n\tilde\nabla_j\tilde\triangle_j) \phi^n_j=0 \eqno (1.1)$$
where the field is defined in the grid points of the (1+1)-dimensional lattice ${\rm \phi
}_{j}^{n}\equiv \phi \left({j\varepsilon, n\tau }\right)$, $\varepsilon, \tau$ being the
space and time fundamental intervals, $j, n$ integer numbers and 
$$\displaylines{
\Delta f_j \equiv f_{j+1}-f_j \quad ,\quad\nabla f_j=f_j-f_{j-1} \cr
\tilde\Delta f_j \equiv {1 \over 2}(f_{j+1}+f_j) \quad , \quad \tilde\nabla f_j
\equiv {1 \over 2}(f_j+f_{j-1}) \cr} $$  
and similarly for the time index. 

Using the method of separation of variables it can easily be proved that the following functions
of discrete variables are solutions of the wave equation (1.1):
$${f}_{j}^{n}\left({k,\omega }\right)={\left({{1+{1 \over 2}i\varepsilon k \over 1-{1 \over
2}i\varepsilon k}}\right)}^{j}{\left({{1-{1 \over 2}i\tau \omega  \over 1+{1 \over 2}i\tau \omega
}}\right)}^{n} \eqno (1.2)$$
\hbox to 16.5 truecm{provided the ``dispersion relationÓ is satisfied: \qquad $\omega^2-k^2=M^2$
\hfill (1.3)}
\noindent{$M$ being the mass of the particle.}

In the limit, $j \rightarrow \infty$, $n \rightarrow \infty$, $j\varepsilon \rightarrow x$, $n\tau
\rightarrow t$ the functions (1.2) becomes plane wave solutions
$$f^n_j(k,\omega) \rightarrow exp\ i(kx-\omega t) \eqno (1.4)$$

Imposing boundary conditions on the space indices
$$f^n_0(k,\omega) = f^n_N(k,\omega) \eqno (1.5)$$
\hbox to 16.5 truecm {we get \qquad $k_m = {2 \over \varepsilon}tg {{\pi m} \over N} \qquad,
\qquad m=0, 1, \ldots, N-1 $ \hfill (1.6)}

\hbox to 16.5 truecm{\hskip 16 pt For the positive energy solutions we define \qquad $\omega_m =
+(k^2_m + M^2)^{1 \over 2}$ \hfill (1.7)}

For $n=0$ the functions $f_j(k_m)$ satisfy the orthogonality relations
$${1 \over N}\sum_{j=0}^{N-1}{f_j^*}(k_m)f_j(k_{m'})=\delta_{mm'} \eqno (1.8)$$
For the quantization of the Klein-Gordon field we impose the equal time commutations relations:
$$\eqalignno{
           &[\tilde\nabla_j \tilde\Delta_j \phi^n_j, \tilde\nabla_{j'} \tilde\Delta_{j'}
\pi^n_{j'}]={i \over \varepsilon} \delta_{jj'}  &(1.12)\cr
           &[\tilde\nabla_j \tilde\Delta_j \phi^n_j, \tilde\nabla_{j'} \tilde\Delta_{j'}
\phi^n_{j'}]=0=[\tilde\nabla_j \tilde\Delta_j \pi^n_j,
\tilde\nabla_{k'}\tilde\Delta_{j'}\pi^n_{j'}] &(1.13) \cr }$$
The Hamiltonian for the real scalar field reads:
$${\rm H}_{n}=\varepsilon \sum\nolimits\limits_{j=0}^{N-1} {1 \over
2}\left\{{{\left({{\tilde{\nabla }}_{j}{\tilde{\Delta }}_{j}{\pi }_{j}^{n}}\right)}^{2}-{1 \over
{\varepsilon }^{2}}\left({{\nabla }_{j}{\Delta }_{j}{\phi }_{j}^{n}}\right)\left({{\tilde{\nabla
}}_{j}{\tilde{\Delta }}_{j}{\phi }_{j}^{n}}\right)+{M}^{2}{\left({{\tilde{\nabla
}}_{j}{\tilde{\Delta }}_{j}{\phi }_{j}^{n}}\right)}^{2}}\right\} \eqno (1.14)$$
A convenient scheme for the Heisenberg equations of motion is
$$\eqalignno{
&{\rm 1 \over \tau }{\Delta }_{n}\left({{\tilde{\nabla }}_{j}{\tilde{\Delta }}_{j}{\phi
}_{j}^{n}}\right)={1 \over i}\left[{{\tilde{\Delta }}_{n}\left({{\tilde{\nabla
}}_{j}{\tilde{\Delta }}_{j}{\phi }_{j}^{n}}\right),{H}_{n}}\right] &(1.15) \cr
&{\rm 1 \over \tau }{\Delta }_{n}\left({{\tilde{\nabla }}_{j}{\tilde{\Delta }}_{j}{\pi
}_{j}^{n}}\right)={1 \over i}\left[{{\tilde{\Delta }}_{n}\left({{\tilde{\nabla
}}_{j}{\tilde{\Delta }}_{j}{\pi }_{j}^{n}}\right),{H}_{n}}\right] &(1.16) \cr}$$
Using (1.12) and (1.13) we get 
$$\eqalignno{
&{ 1 \over \tau }{\Delta }_{n}\left({{\tilde{\nabla }}_{j}{\tilde{\Delta }}_{j}{\phi
}_{j}^{n}}\right)={1 \over i}{\tilde{\Delta }}_{n}\left[{{\tilde{\nabla }}_{j}{\tilde{\Delta
}}_{j}{\phi }_{j}^{n},{H}_{n}}\right]={\Delta }_{n}\left({{\tilde{\nabla }}_{j}{\tilde{\Delta
}}_{j}{\pi }_{j}^{n}}\right) &(1.17) \cr
&{1 \over \tau }{\Delta }_{n}\left({{\tilde{\nabla }}_{j}{\tilde{\Delta }}_{j}{\pi
}_{j}^{n}}\right)={\Delta }_{n}\left\{{{1 \over {\varepsilon }^{2}}{\nabla }_{j}{\Delta
}_{j}{\phi }_{j}^{n}-{M}^{2}{\tilde{\nabla }}_{j}{\tilde{\Delta }}_{j}{\phi }_{j}^{n}}\right\}
&(1.18) \cr}$$
leading to the wave equation (1.1) for the fields $\phi^n_j$ and $\pi^n_j$.

Since the plane wave solutions $f_j^n(k_m,\omega_m), \quad (m=0,1,\ldots,N-1)$ form a complete
set of orthogonal functions with respect to the space indices, we can expand the wave field and
its conjugate momentum as 
$$\eqalignno{
&{\rm \phi }_{j}^{n}={1 \over \sqrt {N\varepsilon }}\sum\nolimits\limits_{m=-N/2}^{N/2-1}
{\left({1+{1 \over 4}{\varepsilon }^{2}{k}_{m}}\right) \over \sqrt {{2\omega
}_{m}}}\left({{a}_{m}{f}_{j}^{n}\left({{k}_{m},{\omega }_{m}}\right)+{a}_{m}^{\dagger
}{f}_{j}^{{}^*n}\left({{k}_{m},{\omega }_{m}}\right)}\right) &(1.19) \cr
&{\rm \pi }_{j}^{n}={-i \over \sqrt {N\varepsilon }}\sum\nolimits\limits_{m=-N/2}^{N/2-1} \sqrt
{{{\omega }_{m} \over 2}}\left({1+{1 \over 4}{\varepsilon
}^{2}{k}_{m}}\right)\left({{a}_{m}{f}_{j}^{n}\left({{k}_{m},{\omega
}_{m}}\right)-{a}_{m}^{\dagger }{f}_{j}^{{}^*n}\left({{k}_{m},{\omega }_{m}}\right)}\right)
&(1.20) \cr}$$ 
Inverting this equations we can calculate with the help of (1.12) and (1.13):
$$[a_m, a^\dagger_{m'}]=\delta_{mm'} \quad , \quad [a_m, a_{m'}]=0 \quad , \quad [a^\dagger_m,
a^\dagger_{m'}]=0 \eqno (1.21)$$
Substituting (1.19) and (1.20) in (1.14) we find
$$\rm H={1 \over 2}\sum\nolimits\limits_{m=-N/2}^{N/2-1} {\omega
}_{m}\left({{a}_{m}{a}_{m}^{\dagger }+{a}_{m}^{\dagger }{a}_{m}}\right) \eqno (1.22)$$
If we define the linear momentum on the lattice as 
$$ P= -\sum\nolimits\limits_{j=0}^{N-1} \left({{\tilde{\nabla
}}_{j}{\tilde{\Delta }}_{j}{\pi }_{j}^{n}}\right)\left({{\tilde{\nabla }}_{j}{\Delta }_{j}{\phi
}_{j}^{n}}\right) \eqno (1.23)$$
we find \quad $ P=\sum\nolimits\limits_{m=-N/2}^{N/2-1} {k}_{m}\left({{a}_{m}{a}_{m}^{\dagger
}+{a}_{m}^{\dagger }{a}_{m}}\right)=\sum\nolimits\limits_{m=-N/2}^{N/2-1} {k}_{m}{a}_{m}^{\dagger
}{a}_{m} $ \quad because the zero point momentum vanish by cancellation of $k_m$ with $k_{-m}$

\vskip 1truecm {\noindent \bf 2. Quantization of the Dirac field} 

\medskip The Hamiltonian for the Dirac field $\psi^n_{\alpha j}$ and its hermitian adjoint
$\psi^{\dagger n}_{\alpha j}$ can be defined as: 
$${\rm H}_{n}=\varepsilon \sum\nolimits\limits_{j=0}^{N-1} {{\tilde{\Delta }}_{j}
{\psi }_{j}^{\dagger n}\left\{{{\gamma }_{4}{\gamma }_{1}{1 \over {\varepsilon }^{}}{\Delta
}_{j}{\psi }_{j}^{n}+M{\gamma }_{4}{\tilde{\Delta }}_{j}{\psi }_{j}^{n}}\right\}} \eqno (2.1)$$
\hbox to 16.5 truecm {with \quad ${\rm \gamma }_{1}=\left({\matrix{0&-i\cr
i&0\cr}}\right),{\gamma }_{4}=\left({\matrix{1&0\cr
0&-1\cr}}\right),i{\gamma }_{1}{\gamma }_{4}={\gamma }_{5}=\left({\matrix{0&-1\cr
-1&0\cr}}\right) $ \hfill (2.2)}

We impose the equal time anticommutation relations for the fields $\psi^n_{\alpha j}$ and
$\psi^{\dagger n}_{\alpha j}$
$$\eqalignno{
&{{\left[{{\tilde{\Delta }}_{j}{\psi }_{\alpha j}^{n},{\tilde{\Delta }}_{j'}{\psi }_{\beta
j'}^{\dagger n}}\right]}_{+}={1 \over {\varepsilon }^{}}{\delta }_{\alpha \beta }{\delta
}_{jj'}} &(2.3) \cr
&{ {\left[{{\tilde{\Delta }}_{j}{\psi }_{\alpha j}^{n},{\tilde{\Delta
}}_{j'}{\psi }_{\beta j'}^{n}}\right]}_{+}={\left[{{\tilde{\Delta }}_{j}{\psi }_{\alpha j}^{\dag
n},{\tilde{\Delta }}_{j'}{\psi }_{\beta j'}^{\dagger n}}\right]}_{+}=0} &(2.4) \cr}$$
Using these relations we obtain for the Heisenberg equation of motion 
$$\eqalignno{
{1 \over \tau }{\Delta }_{n}\left({{\tilde{\Delta }}_{j}{\psi }_{j}^{n}}\right) &={1\over
i}\left[{{\tilde{\Delta }}_{n}\left({{\tilde{\Delta }}_{j}{\psi
}_{j}^{n}}\right),{H}_{n}}\right]={1 \over i}{\tilde{\Delta }}_{n}\left[{{\tilde{\Delta}}_{j}{\psi
}_{j}^{n},{H}_{n}}\right] \cr
&={1 \over i}{\tilde{\Delta }}_{n}\left\{{{\gamma }_{4}{\gamma }_{1}{1
\over \varepsilon }{\Delta }_{j}{\psi }_{j}^{n}+M{\gamma }_{4}{\tilde{\Delta }}_{j}{\psi
}_{j}^{n}}\right\} &(2.5)\cr }$$ from which it can be derived the Dirac equation on the lattice:
$$\left({{\rm \gamma }_{1}{1 \over \varepsilon }{\Delta }_{j}{\tilde{\Delta }}_{n}-{i\gamma }_{4}{1
\over \tau }{\Delta }_{n}{\tilde{\Delta }}_{j}+M{\tilde{\Delta }}_{j}{\tilde{\Delta
}}_{n}}\right){\psi }_{j}^{n}=0 \eqno (2.6)$$ 
Applying the operator ${\rm \gamma }_{1}{1 \over
\varepsilon }{\nabla }_{j}{\tilde{\nabla }}_{n}-{i\gamma }_{4}{1 \over \tau }{\nabla
}_{n}{\tilde{\nabla }}_{j}-M{\tilde{\nabla }}_{j}{\tilde{\nabla }}_{n}$ on both sides of (2.6) we
recover the wave equation (1.1) for $\psi^n_j$.

\vskip 1truecm {\noindent \bf 3. Conservation laws and Lorentz invariance} 

\medskip As in the continuous case we can make the connection between symmetries and conservation
laws in the language of generators. The condition for symmetry under u space and time
displacement and pure Lorentz transformation is that the generators are constant of the motion
. In the case of the Klein-Gordon fields the generators of the (one step) space and time
translations and Lorentz boost  can be taken as: 
$$\eqalignno{
P&=-\sum\nolimits\limits_{j=0}^{N-1} {1 \over 2}\left\{{\left({{\tilde{\nabla
}}_{j}{\tilde{\Delta }}_{j}{\pi }_{j}^{n}}\right)\left({{\tilde{\nabla }}_{j}{\Delta }_{j}{\phi
}_{j}^{n}}\right)+\left({{\tilde{\nabla }}_{j}{\Delta }_{j}{\phi
}_{j}^{n}}\right)\left({{\tilde{\nabla }}_{j}{\tilde{\Delta }}_{j}{\pi
}_{j}^{n}}\right)}\right\} &(3.1) \cr
H&=\varepsilon \sum\nolimits\limits_{j=0}^{N-1} {1 \over 2}\left\{{{\left({{\tilde{\nabla
}}_{j}{\tilde{\Delta }}_{j}{\pi }_{j}^{n}}\right)}^{2}+{1 \over {\varepsilon
}^{2}}{\left({{\tilde{\nabla }}_{j}{\Delta }_{j}{\phi
}_{j}^{n}}\right)}^{2}+{M}^{2}{\left({{\tilde{\nabla }}_{j}{\tilde{\Delta }}_{j}{\phi
}_{j}^{n}}\right)}^{2}}\right\} &(3.2)  \cr
K&=\varepsilon \sum\nolimits\limits_{j=0}^{N-1} {1 \over 2}\varepsilon j
\left\{{{\left({{\tilde{\Delta }}_{j}{\tilde{\nabla }}_{j}{\tilde{\Delta }}_{j}{\pi
}_{j}^{n}}\right)}^{2}+{1 \over {\varepsilon }^{2}}{\left({{\tilde{\Delta }}_{j}{\tilde{\nabla
}}_{j}{{\Delta }_{j}\phi }_{j}^{n}}\right)}^{2}+{M}^{2}{\left({{\tilde{\Delta }}_{j}{\tilde{\nabla
}}_{j}{\tilde{\Delta }}_{j}{\phi }_{j}^{n}}\right)}^{2}}\right\} \cr
&\qquad -\varepsilon \sum\nolimits\limits_{j=0}^{N-1} n\tau {1
\over 2}\left\{{\left({{\tilde{\Delta }}_{j}{{\tilde{\nabla }}_{j}\phi
}_{j}^{n}}\right)\left({{\tilde{\Delta }}_{j}{{\nabla }_{j}\phi
}_{j}^{n}}\right)+\left({{\tilde{\Delta }}_{j}{{\nabla }_{j}\phi
}_{j}^{n}}\right)\left({{\tilde{\Delta }}_{j}{{\tilde{\nabla }}_{j}\pi
}_{j}^{n}}\right)}\right\} &(3.3) \cr}$$
Using (1.17) and (1.18) it can be proved that these operators are constant of time 
$${\rm 1 \over \tau }{\Delta }_{n}P={1 \over \tau }{\Delta }_{n}H={1 \over \tau }{\Delta
}_{n}K=0 \eqno (3.4)$$
In order to check the Lorentz invariance of quantized field scheme on the lattice, one can prove
with the help of (1.12) and (1.13) that these operators satisfy the standard commutation
relations: 
$$[H,P]=0 \qquad , \qquad [K,H]=iP \qquad , \qquad [K,P]=iH \eqno (3.5)$$
For the Dirac quantum fields, the generators of the (one step) space and time translations and
Lorentz boost can be taken as 
$$\eqalignno{
P&=-i\sum\nolimits\limits_{j=0}^{N-1} (\tilde\Delta_j \psi_{\alpha j}^{\dagger
n})(\Delta_j\psi_{\alpha j}^n) &(3.6) \cr 
H&=\varepsilon\sum\nolimits\limits_{j=0}^{N-1}\tilde\Delta_j\psi_{\alpha j}^{\dagger
n}\{{(\gamma_4\gamma_1)}_{\alpha \beta }{1 \over \varepsilon }\Delta_j\psi_{\beta j}^n + M{(\gamma
_4)}_{\alpha \beta} \tilde\Delta_j\psi_{\beta j}^n \} &(3.7) \cr 
M_{14}&=i\varepsilon \sum\nolimits\limits_{j=0}^{N-1} \varepsilon j\{\tilde\Delta
_j\tilde\nabla_j\psi_{\alpha j}^{\dagger n}\{{(\gamma_4\gamma_1)}_{\alpha \beta}{1
\over \varepsilon }\Delta_j\tilde\nabla_j\psi_{\beta j}^n + M(\gamma_4)_{\alpha
\beta}\tilde\Delta_j\tilde\nabla_j\psi_{\beta j}^n\}\} \cr 
&-i\varepsilon\sum\nolimits\limits_{j=0}^{N-1} \tau n\{\tilde\Delta_j\psi_{\alpha
j}^{\dagger n}\Delta_j\psi_{\alpha j}^n \}+\varepsilon
\sum\nolimits\limits_{j=0}^{N-1}\tilde\Delta\psi_{\alpha j}^\dagger{1 \over
2i}{(\gamma_1\gamma_4)}_{\alpha \beta }\tilde\Delta_j\psi_{\beta j}^n &(3.8) \cr }$$ 
Using (2.5), (2,3) and (2,4) it can be proved that these operators are constant of time  
$${1 \over \tau}\Delta_nP={1 \over \tau }\Delta_nH={1 \over \tau }\Delta_nM_{14}=0 \eqno
(3.9)$$ 
and that they satisfy the standard commutation relations  
$$[H,P]=0 \quad , \quad [H,M_{14}]=P \quad , \quad [P,M_{14}]=H \eqno (3.10)$$

\end